\documentclass[]{article}
\usepackage[a4paper, total={6in, 9in}]{geometry}
\usepackage[colorlinks,allcolors=blue]{hyperref}
\usepackage{amsmath}
\title{Emergence of Cosmic Space and Horizon Thermodynamics from Kaniadakis Entropy}
\author{Pranav Prasanthan$^{1}$\footnote{pranavprasanth10@gmail.com}, Sarath Nelleri$^{2}$\footnote{sarathnelleri@gmail.com}, Navaneeth Poonthottathil$^{2}$\footnote{navaneeth@iitk.ac.in} and Sreejith E K$^{1}$\footnote{sreejithekphys@gmail.com} \vspace{0.2cm}  \\ \small $^{1}$Department of Physics, Payyanur College, Payyanur-670327, India \\ \vspace{0.2cm} \small $^{1}$Department of Physics, Indian Institute of Technology, Kanpur-208016, India}
\date{}
\begin{document}
\maketitle

\begin{abstract}
Utilizing Kaniadakis entropy associated with the apparent horizon of the Friedmann-Robertson-Walker (FRW) Universe and applying the emergence of cosmic space paradigm, we deduce the modified Friedmann equation for a non-flat (n+1)-dimensional universe. Employing the first law of thermodynamics, we arrive at the same modified Friedmann equation, showing the connection between emergence of cosmic space and first law of thermodynamics. We also establish the condition to satisfy the Generalized second law of thermodynamics within the Kaniadakis framework. Our study illuminates the intricate connection between the law of emergence and horizon thermodynamics, offering a deeper insight through the lens of Kaniadakis entropy.
\end{abstract}

\section*{Introduction}
\label{sec:intro}
The deep connection between gravity and thermodynamics serves as the driving force behind the emergent understanding of gravity. This intriguing connection was first  realized after the discovery of blackhole thermodynamics by Bekenstein and Hawking\cite{bekenstein1973black,bekenstein1974generalized,hawking1975particle,hawking1976black}. The idea that  gravity and thermodynamics are closely related is supported by evidence from other scenarios including Hooft's investigations into black hole physics and quantum field theory which contributed to the understanding of thermodynamic properties of black holes, establishing the connection between thermodynamics and gravity \cite{hooft1993dimensional} and Susskind's holographic principle which suggests that entropy and other thermodynamic property of a region of spacetime can be described by degrees of freedom on the boundary of that region substantiating the connection \cite{susskind1995world}. Jacobson's derivation of Einstein's field equations from the fundamental Clausius connection on a local Rindler causal horizon leads to a significant advancement in this direction \cite{jacobson1995thermodynamics}. Following this, various schemes were put forward for relating gravity with thermodynamics across a range of gravitational theories \cite{eling2006nonequilibrium,padmanabhan2010thermodynamical}. For instance, Verlinde proposed that gravity emerges as an entropic force resulting from variations in entropy linked to the positions of material objects \cite{verlinde2011origin}. The majority of research on the emergent paradigm treats the gravitational field equations as an emergent phenomenon that emerges from a pre-existing space-time background.

Padmanabhan proposed a new perspective on gravity, viewing spacetime as an emergent structure \cite{padmanabhan2012emergence,padmanabhan2012emergent}. The idea of time emerging from some pre-geometric variables is difficult to comprehend. Treating the space surrounding finite gravitational systems as emergent is equally challenging. Padmanabhan contended that these challenges are resolved within the cosmological framework by selecting the proper time of geodesic observers as the time variable. Thus the spatial expansion of the universe can be described as the emergence of cosmic space with the progress of cosmic time. He could arrive at the Friedmann equation using this new idea in the context of Einstein gravity. Extending this procedure, Cai obtained the Friedmann equations for higher dimensional Gauss-Bonnet, and Lovelock gravities for a spatially flat universe \cite{cai2012emergence}. Following \cite{cai2012emergence}, a successful generalization of Padmanabhan's idea to a non-flat universe was done by Sheykhi in ref.\cite{sheykhi2013friedmann}. By making slight adjustments to Cai's proposal, Sheykhi could derive the Friedmann equations within the frameworks of Einstein, Gauss-Bonnet, and Lovelock gravities for a universe with any spatial curvatures. Following this, Ali arrived at the Friedmann equations regarding a general form of entropy \cite{ali2014emergence}. In references \cite{yang2012emergence,ai2014generalized}, a further generalization to Padmanabhan's proposal was put forth with the assumption that the change in Hubble volume is proportional to a more general function $f(\Delta N,N_{surf})$, where $\Delta N=N_{surf}-N_{bulk}$, rather than altering the degrees of freedom. Eune and Kim introduced an alternative generalization of Padmanabhan's concept, utilizing the proper invariant volume to depict the rate of emergence within a non-flat universe\cite{eune2013emergent}. In Ref.\cite{chang2014friedmann}, Padmanabhan’s conjuncture and its modified versions were extracted from the Friedmann equations. The paper also addresses the challenges involved in extending Padmanabhan's proposal to a non-flat universe. Subsequently, Sheykhi effectively implemented his suggestion in \cite{sheykhi2013emergence} to obtain the modified Friedmann equations in several horizon entropy situations, including Tsallis and Cirto entropy \cite{sheykhi2018modified} and Barrow entropy \cite{sheykhi2021barrow}. Padamanabhans conjecture \cite{padmanabhan2012emergence} has also been used recently to the modified Renyl entropy scenario \cite{komatsu2017cosmological}, Sharma-Mittal entropy scenario \cite{naeem2024correction}, blonic systems\cite{buoninfante2022bekenstein}, minimal length frameworks\cite{ai2014generalized} and brane scenarios \cite{sheykhi2013emergence,sepehri2015emergence,sepehri2016emergence}.

Kaniadakis entropy, which is a generalized form of entropy that accounts for non-extensive statistical mechanics. Here, we aim to explore the effects of Kaniadakis entropy on the cosmological field equations on a higher dimensional spacetime. The effects of the generalized Kaniadakis entropy on the Friedmann equations have been investigated in referrences \cite{lymperis2021modified,moradpour2020generalized,hernandez2022observational,drepanou2022kaniadakis,kumar2024kaniadakis,kumar2023holographic,sheykhi2024corrections}. In ref. \cite{lymperis2021modified}, the authors extracted the modified Friedmann equations for a $3$-dimensional non-flat universe, starting from the relation $-dE=TdS$, where they assumed the apparent horizon radius $\tilde r_A$ to be fixed for an infinitesimal time interval $dt$. New extra terms that constitute an effective dark energy sector depending on the Kaniadakis parameter K were presented in this approach. However, in ref. \cite{sheykhi2024corrections}, Sheykhi deduced the modified Friedmann equations corresponding to the generalized Kaniadakis entropy. The author posited that since entropy is a geometric quantity, any modification on it should  modify the geometric part of the cosmological field equations. So this approach is considered to be more logical than the first. The author also stated that since our universe is expanding, the work term due to the volume change must be included in the first law of thermodynamics, and  modified it as $dE=TdS+WdV$. In ref. \cite{salehi2023accelerating}, Salehi proposed that Kaniadakis entropy plays a crucial role in studying the accelerated expansion of the universe without relying on dark energy. The scope of this paper is to deduce the Friedmann equations in higher dimensional non-flat FRW universe from the law of emergence, and the first law of thermodynamics and to establish the generalized second law of thermodynamics through the lens of Kaniadakis entropy. 

The remaining portion of the paper is organised as follows. The Sect. \ref{sec:2}, provides a brief description about the Kaniadakis statistics. In Sect. \ref{sec:3}, we extract the modified Friedmann equation for $(n+1)$-dimensions from the Kaniadakis entropy employing the law of emergence. In Sect. \ref{sec:4}, we present the thermodynamic derivation of the first and second modified Friedmann equations. In Sect. \ref{sec:5}, we test the validity of the Generalized second law of thermodynamics. Finally, we conclude in  Sect. \ref{sec:6}.\\[5pt]

\section{Kaniadakis statistics}\label{sec:2}

In the realm of cosmology, the advent of Kaniadakis statistics by Giorgio Kaniadakis \cite{kaniadakis2002statistical}, which is a further generlization of Tsallis statistics \cite{tsallis2009introduction} introduces a paradigm shift, offering a fresh perspective on the universe's evolution by incorporating additional parameters, particularly with heavy-tailed distributions. At its core, Kaniadakis statistics extends the principles of standard Boltzmann-Gibbs statistics \cite{tsallis1988possible} by introducing parameters $K-$exponential and $K-$logarithm defined respecitively as
\begin{align}
	exp_K(x)&=(\sqrt{1+K^2x^2}+K x)^{{1}/{K}},\\
	\ln_{\{K\}}x&=\frac{x^K-x^{-K}}{2K}.
\end{align}
Where $K$ is a dimensionless parameter, known as the Kaniadakis parameter which ranges as $-1<K<1$. It signifies the deviation from standard statistical mechanics and in the limit $K\rightarrow 0$, the standard mechanics is restored. Studies on relativistic particle systems \cite{vasyliunas1968survey,hasegawa1985plasma} have indicated a distribution function that deviates from exponential behavior, showing power tails attributed to Kaniadakis entropy \cite{kaniadakis2001non,kaniadakis2002statistical,kaniadakis2005statistical} is given as
\begin{align}\label{ks}
	S_K = k_B \sum_{i} n_i\ln_{\{K\}} n_i,  
\end{align}
where $k_B$ is the Boltzmann constant and $n_i$ is the generalized Boltzmann factor for the $i$-th level of a system given as
\begin{align}\label{tail}
	n_i = \alpha \exp_K[-\beta(E_i - \mu)],
\end{align}
where the parameters takes the form $\alpha = [{(1-K)}/{(1+K)}]^{1/2K}$ and $1/\beta= \sqrt{1-K^2}T$. Also, the Kaniadakis entropy is related to Tsallis entropy in the limit when the non-extensive parameter $q$ of Tsallis entropy approaches $1$, both reducing to classical Boltzmann-Gibbs entropy \cite{abreu2018tsallis}. One key benefit of Kaniadakis entropy is its ability to maintain the fundamental principles of conventional statistical theory while also being able to recover them in the limiting scenario \cite{kaniadakis2002statistical,kaniadakis2005statistical}.
At low energies, the aforementioned $K$-distribution behaves like the typical Boltzmann distribution, but at high energies, it exhibits a power-law tail. Many researchers have been intrigued by the statistical theory \cite{kaniadakis2001non,kaniadakis2001h,kaniadakis2002statistical,kaniadakis2005statistical,kaniadakis2006towards,kaniadakis2009relativistic,kaniadakis2009maximum,kaniadakis2010relativistic,kaniadakis2011power,kaniadakis2012physical} founded on the distribution outlined in Eq.(\ref{tail}).

From  the introduction of $K-$statistics, it has found significant applications in blackhole thermodynamics.
Using the definition of micro-canonical ensemble, it has been argued that, for the case of black holes, the Kaniadakis entropy in Eq.(\ref{ks}), can be written as \cite{moradpour2020generalized,lymperis2021modified}
\begin{align}\label{aaa}
	S_K=\frac{1}{K}\sinh\left(K S_{BH}\right),
\end{align}
where $S_{BH}= A/4 G$ is the standard Bekenstein-Hawking entropy and $A= 4\pi \tilde r_A^2$ is the area of the hubble horizon. From Eq. (\ref{aaa}), the relation connecting Kaniadakis entropy $(S_K)$ and apparent horizon area ($A_{n+1}$) in ($n+1$)-dimensions is 
\begin{align}\label{sk2}
	S_K=\frac{1}{K}\sinh\left(K\frac{A_{n+1}}{4G_{eff}}\right),
\end{align}
where $A_{n+1}= n\Omega_n \tilde r_A^{n-1}$ is the area of the horizon. We extend this idea to cosmology and obtain the Friedmann equation within the framework of law of emergence utilising the Kaniadakis entropy. 

\section{Law of emergence and Kaniadakis entropy}\label{sec:3}
According to law of emergence paradigm introduced by Padmanabhan \cite{padmanabhan2012emergence}, the spatial expansion
of the universe can be interpreted as the emergence of cosmic space ($dV$) with the progress of cosmic time ($dt$). It can be expressed by a simple law
\begin{align}\label{p1}
	\frac{d V}{dt}= L_p^2(N_{surf}-N_{bulk}),
\end{align}
where $N_{surf}$ is the number of degrees of freedom in the boundary of the apparent horizon of the universe and $N_{bulk}$ is the number of degrees of freedom in the bulk, $V$ is the cosmic volume and $t$ is the cosmic time in planck's unit. However, the Eq. (\ref{p1}) failed to derive the Friedmann equations for a non-flat (FRW) universe in alternative gravity theories \cite{cai2012emergence}. The modification of this proposal was done by Sheykhi \cite{sheykhi2013emergence}, who suggested that in a non-flat universe the expansion law should be generalized as 
\begin{align}\label{shey}
	\frac{dV}{dt}=L_p^2 \frac{\tilde r_A}{H^{-1}}(N_{surf}-N_{bulk}).
\end{align}
This suggests that in a non-flat universe, the increase in volume remains directly related to the discrepancy between the degrees of freedom on the apparent horizon and in the bulk. However, the factor of proportionality is not merely a constant; it is determined by the ratio of the apparent horizon to the Hubble radius. In the case of a spatially flat universe where $\tilde r_A = H^{-1}$, the original proposal in Eq. (\ref{p1}) is restored.
Now, our aim is to derive the modified Friedmann equation from the expression presented in (\ref{shey}), when the entropy associated with the apparent horizon get modified to Kaniadakis entropy in ($n+1$)-dimensions. The Kaniadakis entropy is given by Eq. (\ref{sk2}), where the apparent horizon $\tilde r_A$ is \cite{hayward1998unified}          
\begin{align}\label{tilde}
	\tilde r_A = \frac{1}{\sqrt{H^2 + \frac{k}{a^2}}}.
\end{align}
We assume that the $S_K$ is proportional to an effective area $\Tilde{A}$. Hence the effective area can be expressed as
\begin{align}\label{da2}
	\tilde {A} = \frac{4G^{eff}}{K} \sinh\left(\frac{K n \Omega_n \tilde r_A^{n-1}}{4G_{eff}}\right).
\end{align}
Here, we substituted for $ {A}_{n+1} = \dfrac{n\Omega_n \tilde r_A^{n-1}}{4G_{eff}}$ where $\Tilde{r}_A$ is the radius of the apparent horizon and $\Omega_n $ being the volume of unit $n$-sphere. The time progress of effective volume ($\Tilde{V}$) in terms of $\Tilde{r}_A$ and $\Tilde{A}$ is  expressed as
\begin{align}\label{1}
	\frac{d\tilde V}{dt}= \frac{\tilde r_A}{n-1}\frac{d \tilde A}{dt}.
\end{align}
On substituting Eq. (\ref{da2}) in (\ref{1}), we obtain the time progress of effective volume as
\begin{align}\label{dv3}
	\frac{d\tilde V}{dt}= n\Omega_n \tilde r_A^{n-1} \cosh\left(K \frac{n\Omega_n \tilde r_A^{n-1}}{4G_{eff}}\right) \dot{\tilde r}_A . 
\end{align}
Expressing the right-hand side of Eq. (\ref{dv3}) in terms of a total time derivative yields a more convenient form
\begin{align}\label{dv33}
	\frac{d\tilde V}{dt} = -\frac{n\Omega_n \tilde r_A^{n+2}}{2}\frac{d}{dt}\left[\tilde r_A^{-2}cosh\left[K \frac{n\Omega_n \tilde r_A^{n-1}}{4G_{eff}}\right]-\int_{0}^{\tilde r_A}\dfrac{\sinh\left[K\dfrac{n\Omega_n \tilde r_A^{n-1}}{4G_{eff}}\right]}{\tilde r_A^2} d\left[K\frac{n\Omega_n \tilde r_A^{n-1}}{4G_{eff}}\right]\right].
\end{align}
Motivated by Eq. (\ref{dv33}), we propose the degrees of freedom on the apparent horizon in ($n+1$)-dimensions as
\begin{align}\label{N_surf n+1}
	N_{surf}&=\alpha \frac{n\Omega_n }{G_{eff}}\tilde r_A^{n-1}\cosh\left[K\frac{n\Omega_n \tilde r_A^{n-1}}{4G_{eff}}\right]-\alpha  \frac{ n\Omega_n }{G_{eff}}\tilde r_A^{n+1}\int_{0}^{\tilde r_A} \dfrac{\sinh\left[K\dfrac{n\Omega_n \tilde r_A^{n-1}}{4G_{eff}}\right]}{{\tilde r_A^2}} d\left[K\frac{n\Omega_n \tilde r_A^{n-1}}{4G_{eff}}\right],
\end{align}
where $\alpha=\dfrac{n-1}{2(n-2)}$. The bulk Komar energy in the case of ($n+1$)-dimensions \cite{cai2010friedmann}\cite{padmanabhan2004entropy} is given by
\begin{align}\label{N_}
	E_{Komar}=\frac{(n-2)\rho+np}{n-2}V,
\end{align}
where the volume $V=\Omega_n \tilde r_A^n$. Then the bulk degrees of freedom becomes
\begin{align}\label{N_bulk n}
	N_{bulk}=-\frac{2E_{Komar}}{T}=-4\pi\Omega_n\tilde r_A^{n+1}\frac{(n-2)\rho+np}{n-2},
\end{align}
where $T=\dfrac{1}{2\pi\tilde r_A}$ is the Hawking temperature. A negative sign is introduced in the above equation in order to have $N_{bulk}>0$ in the accelerating phase with $(n-2)\rho+np<0$. We also replace $L_p^2$ with $G^{eff}$ and $V$ with $\tilde V$ in the expression (\ref{shey}) and write it down as
\begin{align}\label{law2}
	\alpha\frac{d\tilde V}{dt}=G_{eff} \frac{\tilde r_A}{H^{-1}}(N_{surf}-N_{bulk}).
\end{align}
Substituting Eq. (\ref{da2}), (\ref{N_surf n+1}) and (\ref{N_bulk n}) in Eq. (\ref{law2}) yields,
\begin{align}\label{FRN1}
	\nonumber-\frac{8\pi G_{eff}}{n(n-1)} ((n-2)\rho+np)= \tilde r_A^{-2}\cosh&\left(K\frac{n\Omega_n \tilde r_A^{n-1}}{4G_{eff}}\right)-\tilde r_A^{-3}\dot {\tilde r}_A H^{-1}\cosh\left(K\frac{n\Omega_n \tilde r_A^{n-1}}{4G_{eff}}\right)\\&-\int_{0}^{\tilde r_A} \dfrac{\sinh\left(K\dfrac{n\Omega_n \tilde r_A^{n-1}}{4G_{eff}}\right)}{\tilde r_A^2} d\left(K\frac{n\Omega_n \tilde r_A^{n-1}}{4G_{eff}}\right).
\end{align}
Using the ($n+1$)-dimensional continuity equation $\dot\rho+nH(\rho+p)=0$ and multiplying the\\ above equation with a factor of $2a\dot a$ followed by adding and subtracting the R.H.S with a term\\ $$a^2 \tilde r_A^{-2} \sinh\left(K\dfrac{n\Omega_n \tilde r_A^{n-1}}{4G_{eff}}\right)\dfrac{d}{dt}\left(K\dfrac{n\Omega_n \tilde r_A^{n-1}}{4G_{eff}}\right),$$  Eq. (\ref{FRN1}) takes the form 

\begin{align}
	\frac{16\pi G_{eff}}{n(n-1)}\frac{d}{dt}(\rho a^2)=\frac{d}{dt}&\left[a^2\tilde r_A^{-2}\cosh\left[K\frac{n\Omega_n \tilde r_A^{n-1}}{4G_{eff}}\right]-a^2\int_{0}^{\tilde r_A} \dfrac{\sinh\left[K\dfrac{n\Omega_n \tilde r_A^{n-1}}{4G_{eff}}\right]}{\tilde r_A^2} d\left[K\frac{n\Omega_n \tilde r_A^{n-1}}{4G_{eff}}\right] \right].
\end{align}
Upon integration, we obtain
\begin{align}\label{frn+1}
	\frac{16\pi G_{eff}}{n(n-1)}\rho=\cosh\left[K\frac{n\Omega_n \tilde r_A^{n-1}}{4G_{eff}}\right]\left[H^2+\frac{k}{a^2}\right]-\int_{0}^{\tilde r_A} \dfrac{\sinh\left[K\dfrac{n\Omega_n \tilde r_A^{n-1}}{4G_{eff}}\right]}{\left(H^2+\frac{k}{a^2}\right)^{-1}} d\left[K\frac{n\Omega_n \tilde r_A^{n-1}}{4G_{eff}}\right].
\end{align}
Here, we substituted $\tilde r_A$ from  Eq. (\ref{tilde}). Any integration constant can be taken to the L.H.S of the equation so that the density $\rho$ is modified (but we are keeping the same symbol $\rho$). The Eq. (\ref{frn+1}) is the modified Friedmann equation in a non-flat ($n+1$)-dimensional FRW universe deduced from the Kaniadakis entropy (\ref{sk2}). For $n=3$ ($G_{eff}\rightarrow G$), the  Eq. (\ref{frn+1}) simplifies to 
\begin{align}\label{frnn}
	\frac{8\pi G}{3}\rho=\cosh\left(K\frac{\pi \tilde r_A^2}{G}\right)\left(H^2+\frac{k}{a^2}\right)-\frac{K\pi}{G}\text{shi}\left(K\frac{\pi}{G(H^2+\frac{k}{a^2})}\right) ,
\end{align}
which is consistent with the result obtained in \cite{lymperis2021modified}  using the first law of thermodynamics. Here, shi($x$) is an entire
mathematical odd function of $x$ with no branch discontinuities given by $$\text{shi}(x)=\int_{0}^{x} \dfrac{\sinh x dx}{x}. $$  Also in the limit $K\rightarrow0$, Eq. (\ref{frnn}) reduces to the Friedmann equations derived in the context of Bekenstein-Hawking entropy \cite{sheykhi2013emergence}.

\section{Kaniadakis entropy and first law of thermodynamics}\label{sec:4} 

In this section, we deduce the modified Friedmann equation in non-flat, ($n+1$)-dimensions by employing the first law of thermodynamics. Sheykhi in his work \cite{sheykhi2024corrections} posited that since entropy is a geometric property, any changes to it would influence the geometric nature of the field equations. This lends a greater coherence to this approach than the one followed in ref. \cite{lymperis2021modified}. Furthermore, given the expanding nature of our universe, it becomes imperative to account for the work term arising from changes in volume.\\[5pt]
The associated temperature at the apparent horizon can be defined as \cite{hayward1998unified,hayward1999dynamic}
\begin{align}\label{st}
	T=\frac{\kappa}{2\pi}=-\frac{1}{2\pi\tilde r_A}\biggl(1-\frac{\dot{\tilde r}_A}{2H \tilde r_A}\biggr),
\end{align}
where $\kappa$ is the surface gravity. The work density associated with expansion is defined as \cite{akbar2007thermodynamic}
\begin{align}\label{w1}
	W = \frac{1}{2}(\rho-p).
\end{align}
The work density represents the work performed as a result of the volume change of the universe caused by the variation in the apparent horizon radius.
The first law of thermodynamics at the apparent horizon is assumed to be satisfied, and has the form
\begin{align}\label{f1}
	dE=TdS+WdV.
\end{align}
The energy $E = \rho V$ represents the total energy of the material enclosed within the apparent horizon, which is actually direct ($n+1$) dimensional  generalization of the ($3+1$) dimensional one, given by Misner and Sharp \cite{akbar2007thermodynamic, misner1964relativistic}. The energy ($E$) is different from the Komar energy, ($E_{Komar}$) used in emergence of cosmic space. Here, $V$ is the volume enclosed by $n$-dimensional sphere with area $A_{n+1}=n\Omega_n \tilde r_A^{n-1}$. This law reduces to the conventional law $dE = TdS - pdV$, for a pure de Sitter phase with $\rho=-p$.
Taking the differential of $E$, we get
\begin{align}\label{de}
	dE= \rho n \Omega_n \tilde r_A^{n-1} d\tilde r_A + \Omega_n \tilde r_A^n \dot\rho dt.
\end{align}
Using the continuity equation $\dot\rho + nH(\rho+p)=0$, we obtain
\begin{align}\label{de1}
	dE= \rho n \Omega_n \tilde r_A^{n-1} d\tilde r_A - n\Omega_n \tilde r_A^n(\rho +p)H dt.
\end{align}
We further assume the entropy associated with the apparent horizon is in the form of Kaniadakis entropy (\ref{sk2}). Differentiating the modified entropy-area relation (\ref{sk2}), we get
\begin{align}\label{ds}
	dS=\frac{n(n-1)\Omega_n}{4G_{eff}}\tilde r_A^{n-2} \cosh\left(K\frac{n \Omega_n \tilde r_A^{n-1}}{4G_{eff}}\right) d\tilde r_A.
\end{align}
Substituting Eq. (\ref{st}), (\ref{w1}), (\ref{de1}) and (\ref{ds}) in Eq. (\ref{f1}), we obtain
\begin{align}\label{4}
	(\rho+p)H dt = \frac{(n-1)}{4G_{eff}}\frac{1}{2\pi \tilde r_A^{3}}\cosh\left(K \frac{n\Omega_n \tilde r_A^{n-1}}{4G_{eff}}\right)  d\tilde r_A.
\end{align}
Using the continuity equation $\dot\rho + nH(\rho+p)=0$, the Eq. (\ref{ds}) reduces to 
\begin{align}
	-\frac{8\pi G_{eff}}{n(n-1)} \dot\rho dt =\tilde r_A^{-3}\cosh\left(K \frac{n\Omega_n \tilde r_A^{n-1}}{4G_{eff}}\right)  d\tilde r_A.
\end{align}
Upon integration , we get
\begin{align}\label{t22}
	\frac{16\pi G_{eff}}{n(n-1)}\rho=\cosh\left[K\frac{n\Omega_n \tilde r_A^{n-1}}{4G_{eff}}\right]\left[H^2+\frac{k}{a^2}\right]-\int_{0}^{\tilde r_A} \dfrac{\sinh\left[K\dfrac{n\Omega_n \tilde r_A^{n-1}}{4G_{eff}}\right]}{\left(H^2+\frac{k}{a^2}\right)^{-1}} d\left[K\frac{n\Omega_n \tilde r_A^{n-1}}{4G_{eff}}\right],
\end{align}
which is exactly the modified Friedmann equation presented in Eq. (\ref{frn+1}), obtained from the law of emergence in the section \ref{sec:3}. This suggest that the Kaniadakis entropy is consistent with the connection between the law of emergence and laws of thermodynamics. It is important to highlight the efforts to derive the expansion law from the first law of thermodynamics done in various references \cite{dezaki2015generalized,wang2015cosmological,heydarzade2016emergent,hashemi2015laws,zhang2018quantum,tu2018accelerated,komatsu2019generalized,vt2022emergence,mahith2018expansion}. The revised expansion laws of the universe, as suggested by Cai \cite{cai2012emergence}, Sheykhi \cite{sheykhi2013emergence}, and Yang \cite{yang2012emergence}, can be traced from the the fundamental equation of the first law of thermodynamics, which is expressed as $dE = TdS + WdV$. These studies demonstrate that the first law of thermodynamics can be viewed as the fundamental basis for understanding the expansion law of the universe. 

Combining the continuity equation $\dot \rho + nH (\rho+p)=0$, with the first modified Friedmann equation (\ref{t22}), we can deduce the second modified Friedmann equation for a non-flat ($n+1$) dimensional FRW universe in the context Kaniadakis entropy. Taking the time derivative of the first Friedmann equation (\ref{t22}), we obtain 
\begin{align}\label{t221}
	\frac{8\pi G_{eff}}{n(n-1)}\dot \rho=\cosh\left(K\frac{n\Omega_n \tilde r_A^{n-1}} {4G_{eff}}\right) H \left(\dot H- \frac{k}{a^2}\right).
\end{align}
With the help of continuity equation $\dot \rho + nH (\rho+p)=0$, the Eq. (\ref{t221}) becomes
\begin{align}\label{t2222}
	-\frac{8\pi G_{eff}}{(n-1)}(\rho+p)= \cosh\left(K\frac{n\Omega_n \tilde r_A^{n-1}} {4G_{eff}}\right) \left(\dot H-\frac{k}{a^2}\right).
\end{align}
Now utilizing the relation $\dot H = {\ddot a}/{a}-H^2$ and substituting for $\rho$ from the first Friedmann equation (\ref{t22}), Eq. (\ref{t2222}) rearranges to 
\begin{align}\label{t2333}
	\nonumber-\frac{8\pi G_{eff}}{(n-1)}p- \frac{n}{2}\cosh\left(K\frac{n\Omega_n \tilde r_A^{n-1}} {4G_{eff}}\right)\left[H^2+ \frac{k}{a^2}\right]&-\frac{n}{2}\int_{0}^{\tilde r_A} \dfrac{\sinh\left[K\dfrac{n\Omega_n \tilde r_A^{n-1}}{4G_{eff}}\right]}{\left(H^2+\frac{k}{a^2}\right)^{-1}} d\left[K\frac{n\Omega_n \tilde r_A^{n-1}}{4G_{eff}}\right]\\&=\cosh\left(K\frac{n\Omega_n \tilde r_A^{n-1}} {4G_{eff}}\right) \left(\frac{\ddot a}{a}-H^2-\frac{k}{a^2}\right).
\end{align}
Simplifying and rearranging Eq. (\ref{t2333}), we arrive at
\begin{align}\label{sec22}
	\nonumber\cosh\left[K\frac{n\Omega_n \tilde r_A^{n-1}} {4G_{eff}}\right]\left[2\frac{\ddot a}{a}+(n-2)\left[H^2+\frac{k}{a^2}\right]\right]- n\int_{0}^{\tilde r_A} \dfrac{\sinh\left[K\dfrac{n\Omega_n \tilde r_A^{n-1}}{4G_{eff}}\right]}{\left(H^2+\frac{k}{a^2}\right)^{-1}} d\left[K\frac{n\Omega_n \tilde r_A^{n-1}}{4G_{eff}}\right]\\=-\frac{16 \pi G_{eff}}{(n-1)}p,
\end{align}
which is the second modified Friedmann equation governing the evolution of a non-flat ($n+1$) dimensional FRW universe deduced from the Kaniadakis entropy. For $n=3$ ($G_{eff}\rightarrow G$), the Eq. (\ref{sec22}) simplifies to 
\begin{align}\label{sec45}
	\cosh\left[K\frac{\pi \tilde r_A^2}{G}\right]\left[2 \frac{\ddot a}{a} + \left[H^2+\frac{k}{a^2}\right]\right]- 3\frac{K\pi}{G}\text{shi}\left(K\frac{\pi}{G(H^2+\frac{k}{a^2})}\right)= -8\pi G p,
\end{align}
which is the second modified Friedmann equation for a $3$-dimensional non flat FRW universe based on Kaniadakis entropy. In the limit $K \rightarrow 0$, Eq. (\ref{sec45}) reduces to the second Friedmann equation in standard cosmology
\begin{align}
	2\frac{\ddot a}{a} + H^2+\frac{k}{a^2}= -8\pi G p.
\end{align}
Combining the first and second modified Friedmann equations (\ref{t22}) and (\ref{sec22}), gives the second time derivative of the scale factor as 
\begin{align}\label{sd}
	\frac{\ddot a}{a}\cosh\left[K\frac{n\Omega_n \tilde r_A^{n-1}} {4G_{eff}}\right]  - \int_{0}^{\tilde r_A} \dfrac{\sinh\left[K\dfrac{n\Omega_n \tilde r_A^{n-1}}{4G_{eff}}\right]}{\left(H^2+\frac{k}{a^2}\right)^{-1}} d\left[K\frac{n\Omega_n \tilde r_A^{n-1}}{4G_{eff}}\right]= -\frac{8 \pi G_{eff}}{n(n-1)}[(n-2)\rho + np].
\end{align}
Using the equation of state parameter $\omega = p/\rho$, the Eq. (\ref{sd}) becomes
\begin{align}\label{34}
	\frac{\ddot a}{a}\cosh\left[K\frac{n\Omega_n \tilde r_A^{n-1}} {4G_{eff}}\right]  - \int_{0}^{\tilde r_A} \dfrac{\sinh\left[K\dfrac{n\Omega_n \tilde r_A^{n-1}}{4G_{eff}}\right]}{\left(H^2+\frac{k}{a^2}\right)^{-1}} d\left[K\frac{n\Omega_n \tilde r_A^{n-1}}{4G_{eff}}\right] = -\frac{8 \pi G_{eff}}{n(n-1)} \rho [(n-2)+n \omega].
\end{align}
The condition for the accelerated expansion of the universe ($\ddot a > 0$), requires
\begin{align}
	(n-2)+n\omega < 0 \;  \rightarrow \omega < -\frac{n-2}{n}.
\end{align}
For $n=3$, we obtain the inequality $\omega < -\dfrac{1}{3}$ in Friedmann cosmology. It typically indicates that the cosmic fluid has a negative pressure relative to its energy density. This condition is often associated with certain types of energy components such as dark energy, which exert a repulsive gravitational force, causing the universe's expansion to accelerate.\\ The equations (\ref{t22}) and (\ref{34}) depicts the evolution of a ($n+1$) dimensional FRW universe with any spatial curvature through the lens of Kaniadakis entropy. The cosmological consequences of the above derived Friedmann equations are left for future research.

\section{Generalized Second Law of Thermodynamics}\label{sec:5}
Our aim here is to establish the generalized second law of thermodynamics when the entropy associated with the horizon is given by the Kaniadakis entropy (\ref{sk2}). The authors in Ref. \cite{wang2006thermodynamics,zhou2007generalized,sheykhi2009generalized}, delved into the investigation of generalized second law of thermodynamics within the framework of an accelerating universe. Using Eq. (\ref{4}), solving for $\dot{\tilde r}_A$ gives
\begin{align}\label{vvvvvvv}
	\dot {\tilde r}_A = \frac{8 \pi G_{eff}}{(n-1)} \frac{\tilde r_A^3 H (\rho + p)}{\cosh\left[K\dfrac{n\Omega_n \tilde r_A^{n-1}} {4G_{eff}}\right]}.
\end{align}
When the dominant energy condition holds, $\rho+p\ge 0 $, implies $\dot {\tilde r}_A \ge 0$. Next we calculate $T_h \dot S_h$.
\begin{align}
	T_h \dot S_h = \frac{1}{2\pi \tilde r_A}\left(1-\frac{\dot {\tilde r}_A}{2H\tilde r_A}\right) \frac{d}{dt}\left[\frac{1}{K}\sinh\left(K\frac{n\Omega_n \tilde r_A^{n-1}}{4G_{eff}}\right)\right],
\end{align}
upon simplification, we get
\begin{align}\label{gbn}
	T_h \dot S_h = n\Omega_n \tilde r_A^n H (\rho+p)\left(1-\frac{\dot {\tilde r}_A}{2H\tilde r_A}\right).
\end{align}
For an accelerating universe, the equation of state parameter $\omega =p/\rho < -1$, leads to a potential violation of the dominant energy condition, $\rho+p<0$. Consequently, the second law of thermodynamics, $\dot S \ge 0$ is no longer valid. To address this, the total entropy of universe can be taken as $S=S_h+S_m$, where $S_m$ accounts for the entropy of matter field inside the apparent horizon. Hence, it becomes imperative to examine the time evolution of total entropy, $S$. So, if the generalized second law of thermodynamics holds, the condition $\dot S_h+ \dot S_m \ge 0$, should hold true for the total entropy.
Employing the Gibbs equation \cite{izquierdo2006dark}, we have
\begin{align}\label{gb}
	T_mdS_m = d(\rho V)+ pdV = Vd\rho + (\rho+p)d V.
\end{align}
Here, $T_m$ represents the temperature of matter fields. We propose that the boundary of universe maintains thermal equilibrium with the matter field contained in it. This implies that the temperatures of both components are equal, $T_h=T_m$ \cite{izquierdo2006dark}. If the local equilibrium assumption is not strictly maintained, it would entail observing an energy transfer between the horizon and bulk fluid, which is not deemed physically acceptable. Substituting $V= \Omega_n \tilde r_A^n$  and the continuity equation $\dot \rho +nH(\rho+p)=0$ in the Gibbs equation (\ref{gb}), we obtain
\begin{align}\label{gb2}
	T_h\dot S_m = n\Omega_n \tilde r_A^{n-1}(\rho+p)\dot{\tilde r}_A- n\Omega_n \tilde r_A^n H (\rho+p) 
\end{align}
Now, we obtain the total time evolution of $S_h+S_m$, by combining Equations (\ref{gbn}) and (\ref{gb2}) as 
\begin{align}\label{bg}
	T_h(\dot S_h + \dot S_m)= \frac{n}{2}\Omega_n \tilde r_A^{n-1}(\rho+p)\dot{\tilde r}_A 
\end{align}
Substituting for $\dot {\tilde r}_A$ from Eq. (\ref{vvvvvvv}) in Eq. (\ref{bg}), we obtain
\begin{align}\label{cv}
	T_h(\dot S_h + \dot S_m)= \frac{4\pi G_{eff}}{(n-1)} \frac{n\Omega_n \tilde r_A^{n+2}(\rho+p)^2H}{\cosh\left[K\dfrac{n\Omega_n \tilde r_A^{n-1}} {4G_{eff}}\right]},
\end{align}
which is clearly a non-negative function during the evolution of universe. This is the generalized second law of thermodynamics when the entropy associated with the horizon takes the form of Kaniadakis entropy. For $n=3$ ($G_{eff}\rightarrow G$), the Eq. (\ref{cv}) simplifies to
\begin{align}\label{nb}
	T_h(\dot S_h + \dot S_m)= \frac{8\pi^2 G\tilde r_A^5 (\rho+p)^2 H}{\cosh\left[K\dfrac{\pi \tilde r_A^2}{G}\right]}.
\end{align}
In the limit $K\rightarrow0$, Eq. (\ref{nb}) reduces to the generalized second law of thermodynamics in standard cosmology
\begin{align}
	T_h(\dot S_h + \dot S_m)=8\pi^2 G\tilde r_A^5 (\rho+p)^2 H \ge 0.
\end{align}

\section{Conclusions}\label{sec:6}
Kaniadakis entropy is an extension of the Boltzmann-Gibbs entropy to accommodate systems exhibiting long-range interactions or non-extensive properties. Unlike the Boltzmann-Gibbs entropy, which assumes extensive properties (additivity), Kaniadakis entropy embraces non-extensivity, making it suitable for systems where long-range correlations and non-additive behaviors are present. Kaniadakis cosmology incorporates the Kaniadakis entropy and extends the standard cosmological models by considering non-trivial statistical effects and deviations from conventional thermodynamic behavior, especially in extreme conditions such as those encountered in the early universe or near black hole horizons. The Bekenstein entropy-area relation is modified to Kaniadakis entropy inorder to extend the thermodynamic description of blackholes in scenarios where standard thermodynamics may not be applicable. 

In this study, we successfully implemented Sheykhi's modified law of emergence, to extract the modified Friedmann equation for an ($n+1$)-dimensional, non-flat FRW universe utilizing the Kaniadakis entropy. The emergent space paradigm suggests that the expansion of universe or the emergence of cosmic space is driven by the departure from holographic equipartition or it can be seen as a tendency to satisfy the holographic discrepancy. This implies that the expansion of the universe is propelled by the difference between the degrees of freedom existing on the horizon and those within the bulk. This paradigm shifts the perspective from considering spacetime as a fundamental entity to viewing it as an emergent phenomenon. The integration of Kaniadakis cosmology provides a new perspective on cosmological dynamics within the framework of emergent space paradigms. Further, we employed the first law of thermodynamics $dE=TdS+WdV$ in conjunction with Kaniadakis entropy to deduce the modified Friedmann equation, where we considered the work term related to change in volume due to the expansion of the universe. The  Friedmann equations obtained from both perspectives coincides. This coincidence supports the viability of Padmanabhan’s law of emergence and reveals the intimate relationship of the emergent space paradigm with horizon thermodynamics in Kaniadakis cosmology. Combining the continuity equation and the first modified Friedmann equation, we extracted the second modified Friedmann equation for a ($n+1$)-dimensional, non-flat FRW universe within the Kaniadakis framework. Finally, we established the Generalized second law (GSL) of thermodynamics incorporating the Kaniadakis entropy. Here, we assumed that the boundary of the universe maintains thermal equilibrium with the matter field contained in it. Also, considering the time-dependent behaviour of total entropy of the universe, which is taken as the sum of Kaniadakis horizon entropy and matter entropy, we established the condition for the generalized second law of thermodynamics, $\dot S \ge 0$. In the standard limit ($K\rightarrow0$), our results obtained from the Kaniadakis cosmology reduces to the standard cosmology.

In ref. \cite{salehi2023accelerating}, the author showed that the acceleration of the universe can be modelled without relying on dark energy, within the Kaniadakis framework. Hence, it is important to confront the model with the latest observational probes to constraint the parameter space of the model and to test the evidence of Kaniadakis cosmology over the standard cosmology.

\section*{Acknowledgment}

One of the authors Sarath Nelleri acknowledges Indian Institute
of Technology Kanpur for providing the Institute
postdoctoral fellowship. The authors acknowledge Dr.Vinodkumar Thekkeyil and,  Dr. E.K Sreejith, Department of Physics, Payyanur College, for fruitful discussion on the subject. The authors thank Sreerag Radhakrishnan for suggestions on the manuscript.









\begin{thebibliography}{10}
	
	\bibitem{bekenstein1973black}
	J.~D. Bekenstein, ``Black holes and entropy,'' {\em Phys. Rev. D}, vol.~7,
	no.~8, p.~2333, 1973.
	
	\bibitem{bekenstein1974generalized}
	J.~D. Bekenstein, ``Generalized second law of thermodynamics in black-hole
	physics,'' {\em Phys. Rev. D}, vol.~9, no.~12, p.~3292, 1974.
	
	\bibitem{hawking1975particle}
	S.~W. Hawking, ``Particle creation by black holes,'' {\em Commun. math. phys},
	vol.~43, no.~3, pp.~199--220, 1975.
	
	\bibitem{hawking1976black}
	S.~W. Hawking, ``Black holes and thermodynamics,'' {\em Phys. Rev. D}, vol.~13,
	no.~2, p.~191, 1976.
	
	\bibitem{hooft1993dimensional}
	G.~Hooft, ``Dimensional reduction in quantum gravity,'' {\em arXiv preprint
		gr-qc/9310026}, 1993.
	
	\bibitem{susskind1995world}
	L.~Susskind, ``The world as a hologram,'' {\em J. Math. Phys}, vol.~36, no.~11,
	pp.~6377--6396, 1995.
	
	\bibitem{jacobson1995thermodynamics}
	T.~Jacobson, ``Thermodynamics of spacetime: the einstein equation of state,''
	{\em PRL}, vol.~75, no.~7, p.~1260, 1995.
	
	\bibitem{eling2006nonequilibrium}
	C.~Eling, R.~Guedens, and T.~Jacobson, ``Nonequilibrium thermodynamics of
	spacetime,'' {\em Phys. Rev. Lett.}, vol.~96, no.~12, p.~121301, 2006.
	
	\bibitem{padmanabhan2010thermodynamical}
	T.~Padmanabhan, ``Thermodynamical aspects of gravity: new insights,'' {\em Rep.
		Prog. Phy.}, vol.~73, no.~4, p.~046901, 2010.
	
	\bibitem{verlinde2011origin}
	E.~Verlinde, ``On the origin of gravity and the laws of newton,'' {\em JHEP},
	vol.~2011, no.~4, pp.~1--27, 2011.
	
	\bibitem{padmanabhan2012emergence}
	T.~Padmanabhan, ``Emergence and expansion of cosmic space as due to the quest
	for holographic equipartition,'' {\em arXiv:1206.4916}, 2012.
	
	\bibitem{padmanabhan2012emergent}
	T.~Padmanabhan, ``Emergent perspective of gravity and dark energy,'' {\em Res.
		Astro. Astrophys}, vol.~12, no.~8, p.~891, 2012.
	
	\bibitem{cai2012emergence}
	R.-G. Cai, ``Emergence of space and spacetime dynamics of
	friedmann-robertson-walker universe,'' {\em JHEP}, vol.~2012, no.~11,
	pp.~1--8, 2012.
	
	\bibitem{sheykhi2013friedmann}
	A.~Sheykhi, ``Friedmann equations from emergence of cosmic space,'' {\em Phys.
		Rev. D}, vol.~87, no.~6, p.~061501, 2013.
	
	\bibitem{ali2014emergence}
	A.~F. Ali, ``Emergence of cosmic space and minimal length in quantum gravity,''
	{\em Phys. Lett. B}, vol.~732, pp.~335--342, 2014.
	
	\bibitem{yang2012emergence}
	K.~Yang, Y.-X. Liu, and Y.-Q. Wang, ``Emergence of cosmic space and the
	generalized holographic equipartition,'' {\em Phys.Rev. D}, vol.~86, no.~10,
	p.~104013, 2012.
	
	\bibitem{ai2014generalized}
	W.-Y. Ai, H.~Chen, X.-R. Hu, and J.-B. Deng, ``Generalized holographic
	equipartition for friedmann--robertson--walker universes,'' {\em Gen. Rel.
		Grav}, vol.~46, pp.~1--8, 2014.
	
	\bibitem{eune2013emergent}
	M.~Eune and W.~Kim, ``Emergent friedmann equation from the evolution of cosmic
	space revisited,'' {\em Phys. Rev. D}, vol.~88, no.~6, p.~067303, 2013.
	
	\bibitem{chang2014friedmann}
	E.~Chang-Young and D.~Lee, ``Friedmann equation and the emergence of cosmic
	space,'' {\em JHEP}, vol.~2014, no.~4, pp.~1--11, 2014.
	
	\bibitem{sheykhi2013emergence}
	A.~Sheykhi, M.~Dehghani, and S.~Hosseini, ``Emergence of spacetime dynamics in
	entropy corrected and braneworld models,'' {\em JCAP}, vol.~2013, no.~04,
	p.~038, 2013.
	
	\bibitem{sheykhi2018modified}
	A.~Sheykhi, ``Modified friedmann equations from tsallis entropy,'' {\em Phys
		.Lett. B}, vol.~785, pp.~118--126, 2018.
	
	\bibitem{sheykhi2021barrow}
	A.~Sheykhi, ``Barrow entropy corrections to friedmann equations,'' {\em
		Phys.Rev. D}, vol.~103, no.~12, p.~123503, 2021.
	
	\bibitem{komatsu2017cosmological}
	N.~Komatsu, ``Cosmological model from the holographic equipartition law with a
	modified r{\'e}nyi entropy,'' {\em Eur. Phys. J. C}, vol.~77, no.~4, p.~229,
	2017.
	
	\bibitem{naeem2024correction}
	M.~Naeem and A.~Bibi, ``Correction to the friedmann equation with sharma-mittal
	entropy: A new perspective on cosmology,'' {\em Ann. Phys.}, p.~169618, 2024.
	
	\bibitem{buoninfante2022bekenstein}
	L.~Buoninfante, G.~G. Luciano, L.~Petruzziello, and F.~Scardigli, ``Bekenstein
	bound and uncertainty relations,'' {\em Phys. Lett. B}, vol.~824, p.~136818,
	2022.
	
	\bibitem{sepehri2015emergence}
	A.~Sepehri, F.~Rahaman, A.~Pradhan, and I.~H. Sardar, ``Emergence and expansion
	of cosmic space in bionic system,'' {\em Phys.Lett.B}, vol.~741, pp.~92--96,
	2015.
	
	\bibitem{sepehri2016emergence}
	A.~Sepehri, F.~Rahaman, S.~Capozziello, A.~F. Ali, and A.~Pradhan, ``Emergence
	and oscillation of cosmic space by joining m1-branes,'' {\em Euro.Phys.J. C},
	vol.~76, pp.~1--12, 2016.
	
	\bibitem{lymperis2021modified}
	A.~Lymperis, S.~Basilakos, and E.~N. Saridakis, ``Modified cosmology through
	kaniadakis horizon entropy,'' {\em Eur. Phys. J. C}, vol.~81, no.~11,
	p.~1037, 2021.
	
	\bibitem{moradpour2020generalized}
	H.~Moradpour, A.~Ziaie, and M.~K. Zangeneh, ``Generalized entropies and
	corresponding holographic dark energy models,'' {\em Eur. Phys. J. C},
	vol.~80, no.~8, p.~732, 2020.
	
	\bibitem{hernandez2022observational}
	A.~Hern{\'a}ndez-Almada, G.~Leon, J.~Maga{\~n}a, M.~A. Garc{\'\i}a-Aspeitia,
	V.~Motta, E.~N. Saridakis, K.~Yesmakhanova, and A.~D. Millano,
	``Observational constraints and dynamical analysis of kaniadakis
	horizon-entropy cosmology,'' {\em Mont. Notices. Royal Astro. Soc.},
	vol.~512, no.~4, pp.~5122--5134, 2022.
	
	\bibitem{drepanou2022kaniadakis}
	N.~Drepanou, A.~Lymperis, E.~N. Saridakis, and K.~Yesmakhanova, ``Kaniadakis
	holographic dark energy and cosmology,'' {\em Eur. Phys. J. C}, vol.~82,
	no.~5, p.~449, 2022.
	
	\bibitem{kumar2024kaniadakis}
	P.~S. Kumar, B.~D. Pandey, U.~K. Sharma, {\em et~al.}, ``Kaniadakis agegraphic
	dark energy,'' {\em New Astronomy}, vol.~105, p.~102085, 2024.
	
	\bibitem{kumar2023holographic}
	P.~S. Kumar, B.~D. Pandey, U.~K. Sharma, {\em et~al.}, ``Holographic dark
	energy through kaniadakis entropy in non flat universe,'' {\em Eur. Phys. J.
		C}, vol.~83, no.~2, pp.~1--11, 2023.
	
	\bibitem{sheykhi2024corrections}
	A.~Sheykhi, ``Corrections to friedmann equations inspired by kaniadakis
	entropy,'' {\em Phys. Lett. B}, vol.~850, p.~138495, 2024.
	
	\bibitem{salehi2023accelerating}
	A.~Salehi, ``Accelerating universe in kaniadakis cosmology without need of dark
	energy,'' {\em arXiv:2309.15956}, 2023.
	
	\bibitem{kaniadakis2002statistical}
	G.~Kaniadakis, ``Statistical mechanics in the context of special relativity,''
	{\em Phys. Rev. E}, vol.~66, no.~5, p.~056125, 2002.
	
	\bibitem{tsallis2009introduction}
	C.~Tsallis, {\em Introduction to nonextensive statistical mechanics:
		approaching a complex world}, vol.~1.
	\newblock Springer, 2009.
	
	\bibitem{tsallis1988possible}
	C.~Tsallis, ``Possible generalization of boltzmann-gibbs statistics,'' {\em J.
		Stat. Phys.}, vol.~52, pp.~479--487, 1988.
	
	\bibitem{vasyliunas1968survey}
	V.~M. Vasyliunas, ``A survey of low-energy electrons in the evening sector of
	the magnetosphere with ogo 1 and ogo 3,'' {\em J. Geophys. Res.}, vol.~73,
	no.~9, pp.~2839--2884, 1968.
	
	\bibitem{hasegawa1985plasma}
	A.~Hasegawa, K.~Mima, and M.~Duong-van, ``Plasma distribution function in a
	superthermal radiation field,'' {\em Phys. Rev. Lett.}, vol.~54, no.~24,
	p.~2608, 1985.
	
	\bibitem{kaniadakis2001non}
	G.~Kaniadakis, ``Non-linear kinetics underlying generalized statistics,'' {\em
		Phys. A}, vol.~296, no.~3-4, pp.~405--425, 2001.
	
	\bibitem{kaniadakis2005statistical}
	G.~Kaniadakis, ``Statistical mechanics in the context of special relativity.
	ii.,'' {\em Phys. Rev. E}, vol.~72, no.~3, p.~036108, 2005.
	
	\bibitem{abreu2018tsallis}
	E.~M. Abreu, J.~A. Neto, A.~C. Mendes, and A.~Bonilla, ``Tsallis and kaniadakis
	statistics from a point of view of the holographic equipartition law,'' {\em
		Eur. phys. Lett.}, vol.~121, no.~4, p.~45002, 2018.
	
	\bibitem{kaniadakis2001h}
	G.~Kaniadakis, ``H-theorem and generalized entropies within the framework of
	nonlinear kinetics,'' {\em Phys. Lett. A}, vol.~288, no.~5-6, pp.~283--291,
	2001.
	
	\bibitem{kaniadakis2006towards}
	G.~Kaniadakis, ``Towards a relativistic statistical theory,'' {\em Phys. A},
	vol.~365, no.~1, pp.~17--23, 2006.
	
	\bibitem{kaniadakis2009relativistic}
	G.~Kaniadakis, ``Relativistic entropy and related boltzmann kinetics,'' {\em
		Eur. Phys. J. A}, vol.~40, no.~3, p.~275, 2009.
	
	\bibitem{kaniadakis2009maximum}
	G.~Kaniadakis, ``Maximum entropy principle and power-law tailed
	distributions,'' {\em Eur. Phys. J. B}, vol.~70, pp.~3--13, 2009.
	
	\bibitem{kaniadakis2010relativistic}
	G.~Kaniadakis, ``Relativistic kinetics and power-law--tailed distributions,''
	{\em Eurphys. Lett.}, vol.~92, no.~3, p.~35002, 2010.
	
	\bibitem{kaniadakis2011power}
	G.~Kaniadakis, ``Power-law tailed statistical distributions and lorentz
	transformations,'' {\em Phys. Lett. A}, vol.~375, no.~3, pp.~356--359, 2011.
	
	\bibitem{kaniadakis2012physical}
	G.~Kaniadakis, ``Physical origin of the power-law tailed statistical
	distributions,'' {\em Mod. Phys. Lett. B}, vol.~26, no.~10, p.~1250061, 2012.
	
	\bibitem{hayward1998unified}
	S.~A. Hayward, ``Unified first law of black-hole dynamics and relativistic
	thermodynamics,'' {\em Class. Quant. Grav.}, vol.~15, no.~10, p.~3147, 1998.
	
	\bibitem{cai2010friedmann}
	R.-G. Cai, L.-M. Cao, and N.~Ohta, ``Friedmann equations from entropic force,''
	{\em Phys. Rev. D}, vol.~81, no.~6, p.~061501, 2010.
	
	\bibitem{padmanabhan2004entropy}
	T.~Padmanabhan, ``Entropy of static spacetimes and microscopic density of
	states,'' {\em Class. Quant. Grav}, vol.~21, no.~18, p.~4485, 2004.
	
	\bibitem{hayward1999dynamic}
	S.~A. Hayward, S.~Mukohyama, and M.~Ashworth, ``Dynamic black-hole entropy,''
	{\em Phys. Lett. A}, vol.~256, no.~5-6, pp.~347--350, 1999.
	
	\bibitem{akbar2007thermodynamic}
	M.~Akbar and R.-G. Cai, ``Thermodynamic behavior of the friedmann equation at
	the apparent horizon of the frw universe,'' {\em Phys. Rev. D}, vol.~75,
	no.~8, p.~084003, 2007.
	
	\bibitem{misner1964relativistic}
	C.~W. Misner and D.~H. Sharp, ``Relativistic equations for adiabatic,
	spherically symmetric gravitational collapse,'' {\em Phys. Rev}, vol.~136,
	no.~2B, p.~B571, 1964.
	
	\bibitem{dezaki2015generalized}
	F.~L. Dezaki and B.~Mirza, ``Generalized entropies and the expansion law of the
	universe,'' {\em Gen. Rel. .Grav}, vol.~47, pp.~1--14, 2015.
	
	\bibitem{wang2015cosmological}
	Z.-L. Wang, W.-Y. Ai, H.~Chen, and J.-B. Deng, ``Cosmological model from
	emergence of space,'' {\em Phys. Rev. D}, vol.~92, no.~2, p.~024051, 2015.
	
	\bibitem{heydarzade2016emergent}
	Y.~Heydarzade, H.~Hadi, F.~Darabi, and A.~Sheykhi, ``Emergent universe in the
	braneworld scenario,'' {\em Eur. Phys. J. C}, vol.~76, pp.~1--11, 2016.
	
	\bibitem{hashemi2015laws}
	M.~Hashemi, S.~Jalalzadeh, and S.~Vasheghani~Farahani, ``The laws of
	thermodynamics and information for emergent cosmology,'' {\em Gen. Rel.
		Grav.}, vol.~47, pp.~1--12, 2015.
	
	\bibitem{zhang2018quantum}
	W.~Zhang, X.-M. Kuang, {\em et~al.}, ``The quantum effect on friedmann equation
	in frw universe,'' {\em AHEP}, vol.~2018, 2018.
	
	\bibitem{tu2018accelerated}
	F.-Q. Tu, Y.-X. Chen, B.~Sun, and Y.-C. Yang, ``Accelerated expansion of the
	universe based on emergence of space and thermodynamics of the horizon,''
	{\em Phys. Lett. B}, vol.~784, pp.~411--416, 2018.
	
	\bibitem{komatsu2019generalized}
	N.~Komatsu, ``Generalized thermodynamic constraints on
	holographic-principle-based cosmological scenarios,'' {\em Phys. Rev. D},
	vol.~99, no.~4, p.~043523, 2019.
	
	\bibitem{vt2022emergence}
	H.~B. VT, P.~Krishna, K.~Priyesh, and T.~K. Mathew, ``Emergence of space and
	expansion of universe,'' {\em Class. Quant. Grav.}, vol.~39, no.~11,
	p.~115012, 2022.
	
	\bibitem{mahith2018expansion}
	M.~Mahith, P.~Krishna, and T.~K. Mathew, ``Expansion law from first law of
	thermodynamics,'' {\em JCAP}, vol.~2018, no.~12, p.~042, 2018.
	
	\bibitem{wang2006thermodynamics}
	B.~Wang, Y.~Gong, and E.~Abdalla, ``Thermodynamics of an accelerated expanding
	universe,'' {\em Phys. Rev. D}, vol.~74, no.~8, p.~083520, 2006.
	
	\bibitem{zhou2007generalized}
	J.~Zhou, B.~Wang, Y.~Gong, and E.~Abdalla, ``The generalized second law of
	thermodynamics in the accelerating universe,'' {\em Phys. Lett. B}, vol.~652,
	no.~2-3, pp.~86--91, 2007.
	
	\bibitem{sheykhi2009generalized}
	A.~Sheykhi and B.~Wang, ``Generalized second law of thermodynamics in
	gauss--bonnet braneworld,'' {\em Phys. Lett. B}, vol.~678, no.~5,
	pp.~434--437, 2009.
	
	\bibitem{izquierdo2006dark}
	G.~Izquierdo and D.~Pav{\'o}n, ``Dark energy and the generalized second law,''
	{\em Phys. Lett. B}, vol.~633, no.~4-5, pp.~420--426, 2006.
	
\end{thebibliography}

\end{document}